\documentclass[preprint,journal]{vgtc}            % preprint (journal style)
\renewcommand{\manuscriptnotetxt}{Accepted for presentation at IEEE VIS 2026 and to appear in IEEE TVCG in 2027.\vspace{0.32in}}

\onlineid{0}

\preprinttext{}

\vgtccategory{Research}

\title{An Algorithmic Perspective on Information Visualization}

\author{%
  \authororcid{Wouter Meulemans}{0000-0002-4978-3400}
}

\authorfooter{
  \item W. Meulemans is with TU Eindhoven. E-mail: w.meulemans@tue.nl.
}

\abstract{%
Information visualization is inherently a field that brings together various research domains. Roughly speaking, we may identify two perspectives: the design perspective, revolving around how to ensure that a human can work effectively with the visual representations of data and the tools that offer them, and the algorithmic perspective, focusing on how to automatically create such visual representations.
Munzner's model for visualization design places design choices before algorithmic considerations. It offers predominantly a design perspective; as a consequence, applications of this model may consider the algorithmic perspective as an afterthought, bypassing a step that translates the design into the formalism necessary for algorithmic study. As a result, the design may be entangled with the algorithms used to compute a visualization.
Focusing on layout algorithms, we explore the ramifications of this entanglement: quality often goes undefined and unmeasured, and ad-hoc heuristics tend to be applied, reducing trustworthiness and potentially leading to incorrect conclusions. We look at how we may complement Munzner's model---the design perspective---with a clear model of the algorithmic perspective, to obtain a formal, measured understanding of the interplay between visualizations and the algorithms used to create them. 
Paradoxically, the solution lies in a clearer separation of concerns between design and algorithm.
We argue that this model leads to better comparison between approaches, a more fine-grained understanding of their strengths and weaknesses, and allows for uncovering new opportunities, as to eventually lead to a better understanding of visualization.}


\graphicspath{{figs/}{figures/}{pictures/}{images/}{./}} % where to search for the images

\usepackage{tabu}                      % only used for the table example
\usepackage{booktabs}                  % only used for the table example
\usepackage{lipsum}                    % used to generate placeholder text
\usepackage{mwe}                       % used to generate placeholder figures
\usepackage{ccicons}                   % package to be able to use icons from creative commons

\usepackage{mathptmx}                  % use matching math font

\newenvironment{example}{\smallskip\noindent\textbf{Example.}\itshape\hspace{0.25em}}{\smallskip}

\renewcommand{\paragraph}[1]{\medskip\noindent\textbf{{#1}}}

\begin{document}

\firstsection{Introduction}
\maketitle

Visualization and visual analytics support exploring, understanding and communicating patterns and outliers found in datasets \cite{munzner2014visualization}. 
As a discipline, it necessarily brings together various research fields \cite{chen2017pathways}, ranging from design, perception and cognitive psychology, to human-computer interaction, communication and decision making, to data modeling, analytics and algorithms.
We rely on algorithms to automatically construct visualizations, by selecting labels, placing annotations, ordering symbols, deforming shapes, routing connections, and so on. In visual analytics, algorithms are a necessity for creating high-quality visualizations on the fly; but also for communication purposes, algorithms can alleviate designer effort in constructing and optimizing visualizations. 

The established model for visualization design and research is described by Munzner \cite{munzner2009nested}. 
However, from the algorithmic perspective, this model seems incomplete as it jumps from designing the visual encoding (idiom) to developing algorithms. 
Visual idioms tend to describe the graphical representation (how to read the visualization) on an intuitive level or by describing the patterns that one wishes to see.  As such, visual idioms may not be at the level of mathematical formulation necessary to apply standard algorithm-design practice: without a formal \emph{computational problem}, we cannot (for example) prove correctness, optimality or approximation ratio, nor can we analyze quality objectively through computational experiments. Algorithms research tends to take computational problems as a given and visualization research tends to omit such formal definitions, leading to a fundamental disconnect between these two related research fields that both contribute to the eventual efficacy of visualization.

As a result, visualization research often resorts to use ad-hoc heuristics (based on intuitive notions) that generate a visualization that looks good, when tested on (possibly only few) examples: ``qualitative result image analysis'' downstream validation \cite{munzner2009nested}. 
This practice is severely limiting: it entangles visual idioms with the algorithms used to realize them. Visual idioms can wrongfully be concluded as ineffective, when the algorithm did not achieve a solution of sufficiently high ``quality'' –- yet this quality was never defined or measured. As such, one cannot attribute this necessarily to the algorithm or the visual idiom. 

In short, the algorithmic perspective---a common language and understanding of how to design visualization algorithms and how to establish their efficacy---is absent, rather relying largely on a master-apprentice style of learning algorithmic visualization. 

\paragraph{Contributions and organization.}
Our primary purpose is to introduce a clear algorithmic perspective on information visualization, by advocating the central role of modeling visualization problems for algorithmic treatment, and to discuss its implications on visualization research. Herein, we focus on layout algorithms.

We first discuss the role of algorithms in visualization, and their position in Munzner's nested model in more detail in \autoref{sec:alginvis}.
The main observation is that the nested model does not account for the crucial role of modeling problems for algorithmic treatment, to establish efficacious computational problems. By bringing this modeling step to the foreground, we explicitly mediate between the formalism necessary for algorithms research and other perspectives. Paradoxically, this effectively injects a separation between algorithm and visualization, in order to better integrate the research on both. 

In \autoref{sec:algidiom} we introduce the concept of a \emph{metric idiom}, as a way to capture various facets of visualization quality through constraints and measures. We also relate these to formal computational problems as used in algorithms research.
In \autoref{sec:adequacy} we introduce \emph{adequacy} as a way of framing the relation between the metric idiom and actual visualization quality.
We then consider relations between metric idioms, particularly focusing on proxy models and trade-offs, in \autoref{sec:relations}.

We revisit Munzner's model in \autoref{sec:integratemunzner}, to consider how the algorithmic perspective can be integrated into this research model.
In \autoref{sec:convergent-divergent} we briefly reflect on the typical simplify-generalize process used to solve algorithmic challenges and its seeming opposition to convergent-divergent thinking.

In \autoref{sec:examples} we apply the algorithmic perspective to several examples, illustrating how this perspective is useful in framing research, elucidating interesting challenges and uncovering new opportunities.
In \autoref{sec:implications} we discuss the implications of the algorithmic perspective.

\paragraph{Related work.}
This paper aims to establish a model for algorithmic visualization research, broadly defined as supporting visualization construction through well-defined algorithms and understanding the performance of such algorithms in a broader context. As such, it aims to complement Munzner's model for visualization design \cite{munzner2009nested} and subsequent work such as that by Meyer et al.~\cite{meyer2015nested}: an actionable way to frame the development, application and analysis of algorithms in visualization research and design, supported by quantifiable results.

Behrisch et al.~\cite{behrisch2018quality} discuss the role of quality measures throughout the overall visualization pipeline. They formulate visualization construction in general as an optimization function, separating algorithm from quality measure, as also done in this paper. They position the typical measures for visualization quality between low-level and high-level perception measures, classify quality measures, and distinguish implicit and explicit measures. Casting to the algorithmic perspective, an implicit quality measure is part of the metric idiom, but not the computational problem: ideally, the used (explicit) quality measures are then a good proxy (\autoref{sec:proxy}) for the implicit measure. 

More broadly, this paper adds to theory describing information visualization research and practice; see for example the book edited by Chen et al.~\cite{chen2020foundations}.
Chen and Golan frame information visualization as an information-theoretic process \cite{chen2015what}.
Kindlmann and Scheidegger~\cite{kindlmann2014algebraic} describe an algebraic approach to visualization design, particularly affording attention also to the mapping between data and visualization.
Van Wijk~\cite{wijk2005value} adopts various view points to define the value of visualization. Chen et al.~\cite{chen2017pathways} describe various aspects for a theoretical foundation in visualization: in their concepts, the algorithmic perspective provides a \emph{guideline} to developing and establishing algorithm efficacy, though the \emph{theoretic framework} of a metric idiom.

\section{Algorithms in visualization}
\label{sec:alginvis}

Algorithms play a crucial role in visualization and visual analytics. Roughly speaking, we can distinguish three types of algorithms:
\begin{description}
\item[Data handling] algorithms perform automated analyses, with the purpose of selecting, or enriching the information to display. That is, they are generally about \emph{what} data to show. 
\item[Layout] algorithms decide on coordinates, orderings, symbols, sizes, et cetera, for the various elements that make up a visualization. That is, they are generally about \emph{where and how} to show the (possibly selected and/or enriched) data. 
\item[Rendering] algorithms turn a layout into an actual drawing. These algorithms do not make structural decisions on the visualization, but rather aim to represent the layout as faithfully as possible, given the (typically discrete) nature of an image.
\end{description}

We can draw an analogy here to standard computer graphics, where data handling algorithms decide which objects to show, layout algorithms decide where to place these objects into a scene, and rendering algorithms decide how to transform the scene into an image for display.
Throughout this paper, we use \emph{layout} to generally refer to all choices to be made to fully determine the ``scene'', such that everything is specified to sufficient detail as to require only rendering.

We could distinguish a fourth type, algorithms for interaction. Yet, interaction is often to retrieve additional information, place markings, adapt parameters, et cetera. That is, interactions enrich the data with user information and can be interpreted as ``data handling'' algorithms.

Algorithms specifically for (information) visualization are typically layout algorithms: they decide the core structures and quality aspects of an eventual data graphic, based on principles in visualization. 
That is, in computing a layout, they make implicit or explicit choices that affect visualization quality. While this does not necessarily require learning patterns or structures in the data, such techniques can of course help in making layout decisions.

The above focuses on converting a data set into a visualization in a given visual idiom. Yet, algorithms also can also be applied to automate other parts of visualization construction. For example, specification grammars allow for exploring and making inferences about design spaces \cite{satyanarayan2016vega} and suitable visual idioms can be computed based on constraints and preference scores \cite{moritz2019draco}.

\subsection{Munzner's nested model}
\label{sec:munzner}

\begin{figure}[t]
    \centering
    \includegraphics[alt={Three diagrams labeled (a), (b), (c). Diagram (a) shows four nested rectangles are labeled, from outer rectangle to inner rectangle with, ``domain problem characterization'', ``data/operation abstraction design'', ``encoding/interaction technique design'' and ``algorithm design''. Diagram (b) zooms in on ``algorithm design'' and lists ``threat: slow algorithm'', then ``validate: analyze computation complexity'', followed by ``implement system'', and finally ``validate: measure system time/memory''. Diagram (c) mimicks diagram (a), but injects one rectange labeled ``metric idiom design'' between the rectangles for ``encoding/interaction technique design'' and ``algorithm design''},page=1]{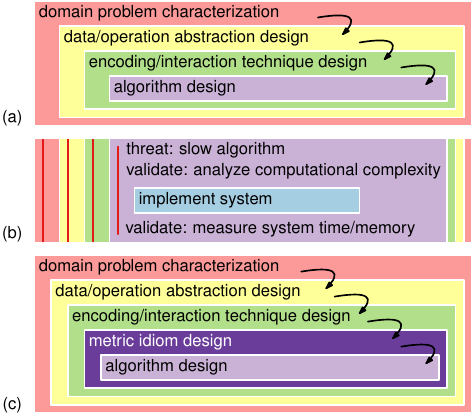}
    \caption{(a) The four nested levels of Munzner's model \cite{munzner2009nested}. (b) The main threats for the algorithmic level in this model \cite{munzner2009nested}. (c) The algorithmic perspective revolves around an intermediary \emph{metric idiom}, which mediates between encoding and algorithm design. Figures (a) and (b) are redrawn with slight changes in colors from \cite{munzner2009nested}.}
    \label{fig:munzner}
\end{figure}

The established model for visualization design and research is described by Munzner \cite{munzner2009nested}. This nested model places visualization research into a four-level hierarchy (see \autoref{fig:munzner}a): 
\begin{description}
\item[Domain problem characterization:] understanding the problem context and what problems the visualization is to help solve.
\item[Data abstraction design:] understanding the data available, abstracting them into general concepts.
\item[Encoding technique design:] obtaining a visual idiom, a description of how to visualize the data.
\item[Algorithm design:] developing an automatic method to construct the visual idiom from data.
\end{description}
Only the last level includes algorithmic consideration, mostly restricted to algorithm efficiency (see \autoref{fig:munzner}b) and a brief consideration of ``correctness'', effectively: does the algorithm generate a representation according to the visual idiom? Indeed, Munzner's nested model affords little attention to algorithm design, rightfully claiming that ``the issues of algorithm design are not unique to visualization, and are extensively discussed in the computer science literature'' \cite{munzner2009nested}.

However, the role that algorithms play in visualization is significantly more fundamental than this model suggests. 
A visual idiom is often at a level considering which objects to show and to a certain degree also how to place them in a layout. But it often does not fully specify the layout.
As such, algorithms make many decisions in finalizing the layout for a visual idiom. Yet, each decision of the algorithm  potentially affects the legibility and interpretation of the eventual visualization. Thus, algorithms have a profound influence on the quality of visualization, which goes far beyond ``correctness'' as used in Munzner’s exposition.

In the nested model, properties of visualization quality are positioned as part of the visual encoding design. This seems fair enough, except that the validation methods do not quite match and it entangles the visual idiom with layout concerns: is a node-link diagram with few edge crossings different (in terms of idiom) from a node-link diagram with right-angle crossings? In the section on flow maps by Munzner \cite{munzner2009nested}, qualitative result image analysis is indicated as the used downstream evaluation for the visual encoding, but is explained as focusing on the algorithm level, rather than the visual encoding itself: effectively a mismatch, but the model leaves comparatively little room for quality considerations at the algorithm level. 

As Munzner also describes \cite{munzner2009nested}, the levels of the nested model could be ``cleaved'' into more levels. The algorithmic perspective complements Munzner's model, giving explicit consideration to modeling visualization quality through metrics. This modeling step effectively injects a new level that was originally implicit in the encoding and algorithmic levels (see \autoref{fig:munzner}c). In \autoref{sec:integratemunzner} we discuss in more detail how the algorithmic perspective integrates with Munzner's nested model. The mismatch described above at least also indicates that this additional level will allow us to better position algorithmic work within the visualization literature.

The subsequent ``nested blocks and guidelines'' model by Meyer et al.~\cite{meyer2015nested} builds on Munzner's nested model, refining concepts into various blocks. This model distinguishes a technique block (roughly, what to visualize) from an algorithm block (how to draw programmatically). They observed that ``Algorithm blocks are often intrinsically connected to technique blocks'' and ``visual encoding techniques are inherently bound to their algorithmic realizations'', which result in ``an inseparable stack of blocks'' \cite{meyer2015nested}. These findings are certainly often (still) true in current practice, and support the importance of an algorithmic perspective in disentangling algorithms from their visual encodings. Indeed, Meyer et al.~\cite{meyer2015nested} state that ``fully characterizing the mappings up from specific algorithms to the visual encodings that they produce remains an important open problem''. Clearly separating \emph{what} to compute from \emph{how} to compute it---the heart of the algorithmic perspective---may provide an answer here, and help practitioners select appropriate algorithms.

\section{The metric idiom}
\label{sec:algidiom}

Central to the algorithmic perspective on visualization is notion that algorithms always solve a computational problem: they optimize \emph{something}. No matter how the algorithm is designed, as soon as it makes decisions on where elements are placed, what colors to use, which elements to show, et cetera, it inherently does so based on aspects of the input for which it is constructing the visualization. The exception perhaps is a fully randomized algorithm. But even a method like random jittering does so to achieve \emph{something}: improved legibility. This observation transcends the nature of the actual visualization problem, the visual idiom and the algorithms used in constructing them. 

\begin{figure}[b]
    \centering
    \includegraphics[alt={Diagram showing four labeled boxes related through colored arrows. A red arrow from the algorithm box to the visualization box is labeled ``generate''; a red arrow from the visualization box to the visual idiom box is labeled ``according to''. A blue arrow from the algorithm box to the computational problem box is labeled ``optimizes''; a blue arrow from the computational problem box is labeled ``captures''. An orange arrow from the computational problem box to the visualization box is labeled ``measures, constrains''. The computational problem box is also labeled with ``(metric idiom)''.},page=1]{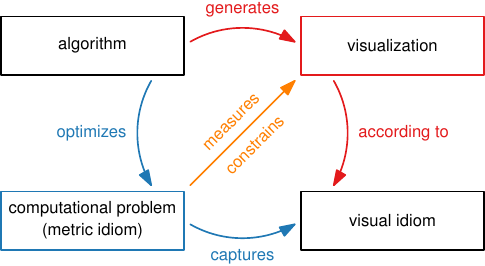}
    \caption{Two ways of relating an algorithm to a visual idiom. In a system, an algorithm generates a visualization according to a visual idiom (red). For designing an algorithm, we need a computational problem that captures quality for the visual idiom (blue). Such a computational problem describes the constraints and quality measure for the eventual visualization, and is typically derived from a multifaceted metric idiom.}
    \label{fig:cps}
\end{figure}

Yet, to implement a visualization tool, one does not need to explicitly think about what this \emph{something} is. The programmer just needs a piece of code---an algorithm---to generate the visualization (its layout) from the data: the red path in \autoref{fig:cps}. 
However, the guidelines, principles and human factors of the design perspective are lost on the computer; it cannot deal with ill-defined problems. Therefore, from an academic point of view, such practice is harmful, as it omits asking \emph{what the problem really is} that the algorithm should solve: the blue path in \autoref{fig:cps}. And therefore, such practices limit the potential for evaluating the quality of methods, and understanding how methods relate or can be applied in other contexts.
If we focus only on \emph{how} to compute, rather than \emph{what} to compute, it becomes hard to evaluate and compare methods. We risk unintentionally showing results selectively, possibly only favorably. Imagine trying to decide between competing products to buy, without specification of purpose, based solely on reviews---likely a limited set and possibly selected for favorable ones.

The root cause here lies with what typically encompasses a visual idiom:
it describes how the data is turned into a visualization, often in the form of how the visualization is to be read, rather than to be constructed. That is, the various choices that an algorithm makes (implicitly or explicitly) are often not captured by a visual idiom. 

\begin{example}
In general, a node-link diagram visualizes a graph by representing each vertex with a symbol (node), and each edge with a curve (link) with its endpoints at the symbols of its incident vertices. But how do we decide where to place the nodes, and how to route the links? The fields of graph drawing and network visualization offer myriad ways on how to address this algorithmic problem, all for the same basic visual idiom. Specific encodings vary in the types of symbols and links permitted, but possibly also prescribe (partially or fully) where to place symbols (such as in geographic settings), or how to route links (for example, by requiring straight line segments only).
\end{example}

\begin{example} 
A scatterplot visualizes elements according to two numeric values by representing each element as a symbol (such as a disk) with its x-coordinate representing one value and its y-coordinate the other value. At a glance, this idiom seems to give little room for algorithmic decisions. Yet, symbol size could be set to reduce clutter, or the drawing order of the symbols could be changed to increase their visibility.
\end{example}

As the above examples show, the choices or optimizations that a visualization algorithm is to make, are not uniquely defined for a visual idiom. There are often multiple ways in which to cast the overarching visualization challenge into a computational one. Hence, it is natural that we need a concept to complement the visual idiom, one that captures visualization quality through formal measures as to allow for algorithmic optimization (the blue path in \autoref{fig:cps}).
We refer to this concept as the \emph{metric idiom}.

A metric idiom aims to capture the visualization design criteria: \emph{what} makes a good visualization? And how can we measure that using objective formulas? This is, by no means, a trivial task as casting the design principles into precise formulas requires a clear understanding of the design perspective for the particular visual idiom.

While visualization quality is often too complex to capture perfectly, we need a metric idiom as to allow for objectively and automatically assessing visualization quality as produced by algorithms. But it should be independent of any algorithm, as to allow for evaluating any layout. Indeed, it should be possible to evaluate, for example, even manual layouts, assuming that they have been encoded appropriately.

The metric idiom serves as an expression of how one aims to measure layout quality.
As such, it should be understood before algorithmic treatment: it is a first step to capturing exactly how one aims to integrate layout quality into the eventual algorithm. 

Below, we discuss the concept of a metric idiom in three steps: \emph{facets} to break down overarching quality into separate notions, \emph{constraints} to describe requirements that a layout is to satisfy in order to meet the basic needs of the idiom, and \emph{measures} that capture quality in the various facets.
A metric idiom is more general than a formal computational problem as used in algorithmic research, but serves as the basis from which one can derive such problems, as discussed in \autoref{sec:computproblem}.

\subsection{Facets}

Often, many different notions contribute to visualization quality \cite{cabouat2026readability}: as quality is multifaceted, we refer to these as \emph{facets} of quality. A facet is to capture a notion that contributes to quality, but this does not have to be at a formal level yet. Our goal here is to break down quality into facets that can be constrained or measured separately. What does it mean for a visualization (in a specific visual idiom) to be ``good''? 

Roughly speaking, there are two types of facets: \emph{legibility} facets that capture how well the various graphical elements in a visualization can been seen; and \emph{correspondence} facets that capture how well the graphical elements and their composition reflect the underlying data and their structures (in line with the Principle of Correspondence~\cite{kindlmann2014algebraic}).

\begin{example}
In a scatter plot, a correspondence facet captures how well the position of an element matches its ``actual'' position according to the two axes. Often, there may be no deviations \cite{cabello2010algorithmic}, and as such is of no consideration to an algorithm. However, various approaches also exist to adjust positions \cite{gartner2025optimizing,giovannangeli2023guaranteed}.
A legibility facet of quality captures how well each of the symbols can be seen, for example, through visible perimeter \cite{cabello2010algorithmic,gartner2025optimizing} or area \cite{giovannangeli2023guaranteed}.
\end{example}

\begin{example}
Cartograms (and other schematic maps) often aim to ``reflect spatial relations well''. However, spatial relations are too general for a facet, as they can interpreted as distance, directions, adjacencies, et cetera. Each of these may be a correspondence facet in the metric idiom. Another correspondence facet is cartographic error, capturing deviation between actual region area and its data value.
Legibility facets focus on how well each region can be seen. Free from overlap is often required. Other legibility facets include for example aspect ratio or shape simplicity, depending on the type of cartogram.
\end{example}

\subsection{Constraints}

The \emph{constraints} of a metric idiom model the hard requirements we set on the layout of a visualization. They describe properties that must be met, otherwise the layout does not achieve a baseline requirement of quality. We call a layout \emph{valid} if it satisfies all constraints, and \emph{invalid} if it violates one or more constraints.

We observe that one can place various constraints on the visualization that are not of immediate concern to a layout algorithm: they reflect parts of the visual design itself, rather that a layout.
To clarify the issue, we distinguish between two types of constraints:
\begin{description}
\item[Layout] constraints reflect parts of the visualization that an eventual algorithm may change: they describe what types of layouts are to be computed. That is, these constraints require explicit consideration such that the algorithm does not violate them. 
\item[Design] constraints reflect parts of the visualization that are derived from the visual idiom or fixed in rendering the layout into the final graphic. These constraints cannot be violated in a layout.
\end{description}

\begin{example} 
For a node-link diagram, we establish the following constraints: (1) all vertices are drawn as circles of radius 1; (2) the circles for vertices may not overlap; (3) all edges are drawn as straight segments between their endpoints.
There are no choices to make algorithmically for constraints (1) and (3): these are design constraints, and refine which variant of node-link diagrams we intend to use. However, the algorithm may choose vertex positions and as such is at risk of making a decision that violates constraint (2): this is a layout constraint.
\end{example}

Whereas layout constraints require attention when designing algorithms, design constraints are practically a given. No matter the choices made by an algorithm, they will be satisfied by design or by appropriate rendering. The design constraints can be thought of as part of the chosen visual idiom and choices made therein for the visual encoding.
Therefore, design constraints have no place in a metric idiom, which should consist only of layout constraints. In the remainder, we thus refer to these simply as constraints.

\subsection{Measures}

Typically, the constraints permit for a variety of valid layouts, and hence we need a way to make quality judgments to compare them and find the ``best'' layout.
The \emph{measures} of a metric idiom capture quality for the purpose of comparing valid layouts. 
A measure maps the layout to a numeric quantity\footnote{Technically, we just need to map visualizations to a (partially) ordered set to allow for comparisons. However, in practice, numerical mappings are typically used and also allow us to reason about difference in quality, approximation guarantees, and so forth.} that reflects the quality of the layout according to that measure. 
For our exposition, we assume the measures reflect \emph{quality} (higher is better) rather than \emph{discrepancy}, distortion or deviation (lower is better).

In general, an idiom may describe multiple measures according to which a visualization is to be assessed in quality, especially for multifaceted problems. Let $(m_1, m_2, \ldots)$ denote these measures, We can then define a partial order $\prec$ on the valid layouts, where $a \prec b$ iff $m_i(a) \leq m_i(b)$ for all $i$ and $m_i(a) < m_i(b)$ for at least one $i$. A layout $M$ is of maximal quality, if there is no other layout $L$ for which $M \prec L$. Note that, with multiple measures, there can be multiple visualizations that are maximal but achieve different levels of quality in the different measures. We revisit this concept in \autoref{sec:tradeoffs}.

\subsection{Computational problem}\label{sec:computproblem}

A metric idiom aims to describe how one wishes to evaluate and estimate eventual visualization quality. As such, it is generally multifaceted, with a variety of measures to nuance evaluation.
However, algorithms can inherently optimize only a single function: when two measures agree the distinction is irrelevant, and if two measures disagree, the algorithm must somehow make a choice. 
Thus, for developing algorithms, we need to cast the metric idiom into a \emph{computational problem}, as used in algorithms research to describe the input assumptions and output requirements for an algorithmic challenge. 

We purposefully omit assumptions from a metric idiom. Assumptions on the data (if any) should readily arise from the visual idiom, application or domain. Otherwise, the metric idiom would not be able to generally assess the quality of visualizations for which it was designed. Further assumptions on the input do allow us to scope the applicability of algorithms. For example, algorithms may require the input graph to be planar, or for coordinates to be distinct. 

The output requirements consist of constraints and a single optimization objective. While the constraints of a metric idiom readily carry over, if an idiom has multiple measures of interest, they need to be narrowed down to a single one. There are, in general, several strategies:
\begin{description}
    \item[Thresholding:] We can turn any number of measures into constraints, by setting a minimum (for quality measures) or maximum value (for discrepancy measures).
    \item[Combining:] We can use sums, products or other functions to combine measures into a single number.
    \item[Prioritizing:] We can aim to find the layout maximizing a second measure, within those layouts that maximize a first measure.
    \item[Omission:] We can simply drop measures from explicit consideration. Especially if they are correlated to other measures, this may be a simple approach. We expand on this idea in \autoref{sec:proxy}.
\end{description}
Depending on the eventual algorithm, one strategy may be easier to work with than another. For example, if the eventual algorithm is a linear program, the sum of measures may be an easy solution, though one must ensure that the sum is meaningful.
As a result of deriving a computational problem from the metric idiom, we have effectively turned the partial order discussed above into a full order. 

Summarizing, a computational problem can be thought of as a metric idiom with (at most) one measure, and possibly with assumptions on the input data. For example, a metric idiom for node-link visualizations of graphs may include measures such as crossings, bends, and resolution. Yet, a specific computational problem can be to compute a layout with a minimal number of bends (the chosen measure) with no crossings (a constraint) for a planar graph (an input assumption).

\subsection{Further considerations}

In developing a metric idiom, we could also consider the computational complexity of actually evaluating a layout. From a practical point of view, being able to efficiently evaluate a metric helps to run experiments. However, our purpose is not to be able to evaluate efficiently, but rather to generate efficiently while understanding the quality obtained. Thus, higher complexity evaluation is not necessarily something to be shied away from. Though, if incorporated directly into the computational problem, the resulting problem is likely computationally complex as well. At the very least, it may be desirable that evaluating a measure is in NP, such that we can determine the actual value, given sufficient resources.
For example, we can compute a reasonably short traveling salesperson tour without knowing the optimal tour \cite{cormen2022introduction}. But to quantify how well an approximation or heuristic does in practice, we may need to be able to find the actual optimum nonetheless.

We should keep in mind that our goal is to compute a layout: it is generally desirable for a computational problem to always permit at least one valid layout for any dataset. We should be careful to avoid overconstraining the problem, already when setting constraints in the metric idiom, but even more so when setting thresholds. After all, while the algorithmic challenge may be interesting, an eventual visualization system would become unreliable if its underlying algorithms would simply conclude that ``no layout exists''.

A completely formal computational problem in algorithms research is to ensure that we can reason about correctness or other quality guarantees. Yet, also for heuristic algorithms, where no such proofs are to be had, such precision is valuable. The metric idiom outlines in broad strokes the different ways in which one could capture facets of quality, yet does not aim to make judgments between facets. By deriving a formal problem, one is forced to make choices on how to deal with competing criteria. As such, it crystallizes decisions made to create the heuristic, and allows for comparison between different heuristics also on a conceptual level: do they make the same or different choices? And downstream, how do these choices manifest in the eventual quality of a visualization?

The multifaceted nature of a metric idiom also mirrors the general idea that different layouts may be suitable for different tasks: there is likely not a single layout that performs better for all tasks and trade-offs need to be made to support a variety of tasks. A computational problem can be thought of as a metric idiom with but a single facet for optimization: in our exposition, its purpose is predominantly to allow comparison of algorithms and application of standard algorithm techniques. In the remainder, we thus mostly use metric idiom, while often simplifying its quality judgment into a single dimension.

\section{Adequacy and solvability}
\label{sec:adequacy}

The purpose of a metric idiom is that we may now think separately about how well quality measures capture visualization quality, and how well an algorithm solves the computational problem.

\begin{figure}[b]
    \centering
    \includegraphics[alt={Diagram of three boxes labeled ``visualization'', ``metric idiom'' and ``human''. An arrow from the metric idiom box to the visualization box is labeled ``measures, constrains''; an arrow from the human box to the visualization box is labeled ``performance, preference''. A line connecting these two arrows is labeled ``adequacy''.},page=1]{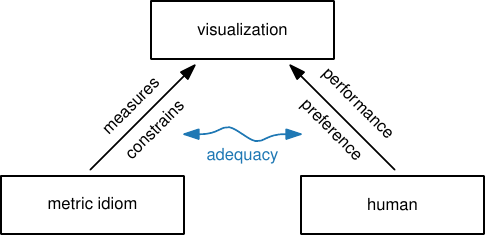}
    \caption{Adequacy refers to the correspondence between visualization quality as established by a metric idiom and as measured via human performance on tasks or preferences. Diagram inspired by a variant used by Helen C. Purchase.}
    \label{fig:adequacy}
\end{figure}

\paragraph{Adequacy.}
A metric idiom aims to predict the quality of a visualization, in terms of task performance or preference by human observers or users. \emph{Adequacy} refers to the correspondence between the metric idiom and human aspects of establishing visualization quality; see also \autoref{fig:adequacy}. A metric idiom is adequate if it predicts the human aspect of visualization quality: if the idiom considers one visualization better than another, then a human should prefer or perform tasks better with the former visualization. The term draws an analogy to statistics, where it indicates absence of significant deviations from (the assumptions of) a model. 

\begin{figure}[t]
    \centering
    \includegraphics[alt={Two diagrams next to each other. The left diagram is a scatter plot, with horizontal axis ``metric idiom'' and vertical axis ``human''. The dots in the bottom left corner are black, and are roughly on a line; the dots in the top right corner are blue and are roughly on this same line. The dots in the top left corner are orange and in the bottom right corner are red. The right diagram is constructed similarly, but the area has been subdivided into four quadrants, labeled roman numeral ``I'' (top left), ``H'' (topright), ``L'' (bottom left) and roman numeral ``II'' (bottom right).},page=1]{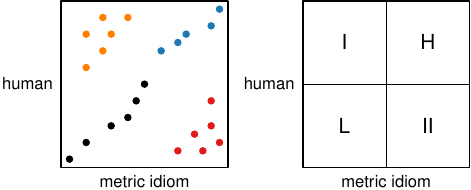}
    \caption{(left) Adequacy as the relation between assessment by humans and by a metric idiom. A dot abstractly represents a visualization. The orange and red dots represent errors or counterexamples to adequacy. The blue and black dots represent good correlation: high adequacy. (right) Interpreting adequacy as hypothesis testing. In case of adequacy, visualizations have high (H) or low (L) performance. Inadequacy can be seen as type-I and type-II errors: the metric idiom respectively underestimates or overestimates quality.}
    \label{fig:hypothesistesting}
\end{figure}

Conceptually, we can aim to relate actual visualization quality (as judged or used by a human) to quality according to a metric idiom; see \autoref{fig:hypothesistesting}. Ideally, we see a perfect positive correlation: high adequacy. Yet, a metric idiom may predict incorrectly if its measures are low, yet the resulting visualization performs well, or if its measures are high, yet the resulting visualization performs poorly. We can draw an analogy here to standard hypothesis testing in statistics. Interpreting a visualization as a hypothesis of adequacy, the former misprediction is a type-I error (false positive), the latter a type-II error (false negative). 

In developing algorithms, we generally aim to achieve high-quality visualizations, that is, in the upper right quadrants in \autoref{fig:hypothesistesting}. While correspondence on low-quality maps does support adequacy to some degree, actually constructing poor visualizations is typically not our goal. But also note that ``high'' and ``low'' are always with respect to alternatives: for large, complex data, measures and performance may be considerably worse than for simple, small data. Nonetheless, the former may be good for what is actually achievable: it can still be a (close to) optimal layout for the given dataset.

We should keep in mind here, that ``quality'' is hard to define absolutely and universally. Preferences and performance may vary between humans, and furthermore depend on the specific task that is considered. That is, a metric idiom is unlikely to be perfectly adequate: the saying ``All models are wrong, some are useful'' \footnote{Commonly attributed to statistician George Box; see also \url{https://en.wikipedia.org/wiki/All_models_are_wrong}, accessed March 2026.} comes to mind. We cannot expect perfect models and we should rather think about the level of adequacy (or fitness for purpose\footnote{See also \url{http://www-stat.wharton.upenn.edu/~steele/Rants/ModelsMandLE.html}, accessed March 2026.}).

In the end, establishing adequacy aims at relating metric predictions to human performance. As such, it is typically done through quantitative evaluation using human subjects: see, for example, \cite{cowley2022framework,lehmann2015study,purchase1995validating,seldmair2015datadriven}. Studying such relations can subsequently lead to novel measures and algorithms (for example, see \cite{zhao2020preserving}).

\begin{figure}[b]
    \centering
    \includegraphics[alt={A scatter plot with horizontal axis ``solvability'' and vertical axis ``adequacy''. It shows four points. 
    Point a is in the bottom left; point b is to the right and above a; point c is to the right of b, and vertically between a and b. A red arrow goes straight down from point b to a point b', which is positioned below all other points.},page=1]{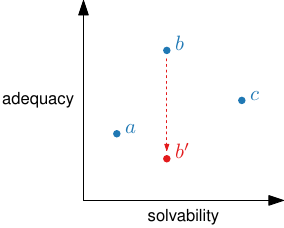}
    \caption{The need to balance between solvability and adequacy. Though both are multifaceted, they are presented here as one dimension for simplicity. Computational problems $b$ and $c$ outperform $a$, as they admit better algorithms while also having higher adequacy. Due to the low solvability, an actual algorithm $b'$ for problem $b$ may achieve much lower adequacy as it cannot get high-quality solutions.}
    \label{fig:adequacytradeoff}
\end{figure}

\paragraph{Solvability.}
A metric idiom lays the foundation for stating computational problems, and thereby eventually aims to support algorithm development. The \emph{solvability} of a computational problem (and thereby indirectly for a metric idiom) is hence the solution quality and resource efficiency of algorithms that aim to solve it. 

Capturing all facets of design---in so far possible---runs the rather probable risk of creating a computational problem that is too complex, thereby defying good algorithmic solutions. On the other hand, while simple problems are easier to develop algorithms for, they may have reduced adequacy. Only in exceptional cases, should we expect that formulas used in automatically assessing quality are anything but proxies for something far more complex.
We thus need to balance between the need for adequacy and solvability; see also \autoref{fig:adequacytradeoff} and \autoref{sec:overlapremoval}.
As such, the metric idiom plays a mediating role between the design perspective and the formalism necessary in the algorithmic perspective.

Adequacy reflects how well measures predict human performance or preference. A high level of adequacy therefore indicates that it is desirable to optimize these measures. Yet, if a computational problem (or indirectly, the metric idiom) has low solvability, an algorithm may not be able to achieve good solutions according to these measures. The algorithm itself may thus exhibit lower adequacy than the idiom it aims to solve; see \autoref{fig:adequacytradeoff}. That is, we may not be able to leverage the high adequacy of a computational problem or metric idiom, if we cannot compute high-quality solutions efficiently enough.

\section{Relating metric idioms and computational problems}
\label{sec:relations}

In bringing the modeling steps necessary for algorithmic treatment to the foreground, the algorithmic perspective allows us to explicitly reason about and compare metric idioms and computational problems. We first highlight two general concepts: proxy models as a way of capturing positively correlated facets, and trade-offs between negatively correlated facets. Subsequently, we briefly discuss the relation between algorithm, computational problem, and metric idiom.

\subsection{Proxy models}
\label{sec:proxy}

Occam's razor, or the principle of parsimony, suggests that between models making the same predictions, the simpler model is preferred. While identical predictions of quality are perhaps unlikely, if the quality measure of one computational problem strongly and positively correlates to the quality of another computational problem, they are, to a certain extent, interchangeable for the purpose of optimizing layout.

In such cases, we may prefer the use of a simpler computational problem over a more complex one, for algorithmic purposes. The simpler model may have higher solvability, thereby allowing us to get high-quality solutions also in the more complex model thanks to its positive correlation.
We can think of such a simpler problem as acting as a \emph{proxy model} for the more complex computational problem.

\begin{figure}[b]
    \centering
    \includegraphics[alt={A chart with horizontal axis ``measure 1'' and vetical axis ``measure 2''. A curved line goes drom from the measure-2 axis towards the measure-1 axis in a decreasing manner. On the righthand side, this line is almost vertical. Three points are shown: point a lies in the bottom left corner, well away from the line; point b lies on the line, and above and to the right of point a; point c lies slightly below the line and to the right of b, but between a and b vertically.},page=1]{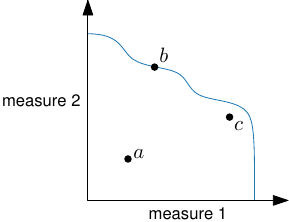}
    \caption{Trade-off between two measures (facets) in a metric idiom. The curve represents the Pareto front of maximal layouts. Layout $a$ is dominated by both other layouts; $b$ and $c$ are incomparable, performing better in one metric but worse in another. Layout $b$ is maximal: it lies on the Pareto front and thus there is no other layout that can perform better in both measures at the same time. Layout $c$ represents a different trade-off to $b$, though it does not lie on the Pareto front: there is a layout that performs better in both measures.}
    \label{fig:tradeoffs}
\end{figure}

\subsection{Trade-offs via multiple measures}
\label{sec:tradeoffs}

A multifaceted metric idiom typically faces trade-offs: performing better in one measure may come at the expense of another. A computational problem must, by necessity, make decisions on this trade-off, by omitting facets, turning them into constraints, or by combining the measures into a single function. 

Though multi-objective optimization is typically difficult to do well, we can nonetheless use algorithms to study the metric idiom and the trade-offs it embodies. There may be various local maxima in the partial order, each modeling a different trade-off between facets. Conceptually, these maxima describe a (discrete or continuous) Pareto front of layouts (\autoref{fig:tradeoffs}).
From its shape, we may learn how the trade-off between facets behaves. In the figure, the steep slope on the right-hand side indicates that we can gain considerably in measure 2, if we are willing to just slightly reduce measure 1. In other words, this front describes a cost-benefit trade-off between the measures.

\begin{figure}[b]
    \centering
    \includegraphics[alt={A chart with horizontal axis ``measure 1'' and vetical axis ``measure 2''. A curved line goes drom from the measure-2 axis towards the measure-1 axis in a decreasing manner. A red point lies on the intersection of this curve and the vertical axis. Halfway through the chart is an orange vertical line: its intersection with the curve is marked with an orange dot. A purple arrow points rightwards and upwards from the origin of the chart, roughly utt not precisely in the direction of a purple dot that lies on the curve.},page=2]{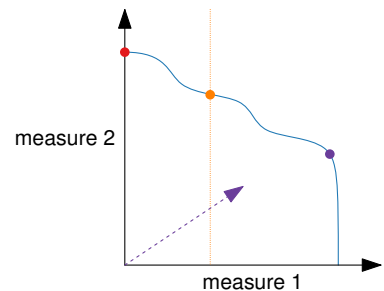}
    \caption{Conceptualization of strategies for reducing the number of measures. Prioritization of measure 2 or omission of measure 1 yields a result that primarily considers measure 2 (red dot): these may behave largely the same, effectively ignoring a facet of the metric idiom. Thresholding gives a lower bound on measure 1 (orange dotted line) and an optimal result (orange dot) must lie to the right of the line. Summing measures yields a direction (purple arrow) and the optimal result then lies on the Pareto point, furthest in the indicated direction (purple dot).}
    \label{fig:strategies}
\end{figure}

Algorithms that tackle computational problems must have made a choice regarding these facets. Indeed, we can interpret the various strategies for reducing the number of measures to one (see \autoref{sec:computproblem}) as how we search the Pareto front for an optimal solution; see \autoref{fig:strategies}.

By explicitly mapping where the layouts produced by algorithms or defined via computational problems lie in this space, we may also explore new options for algorithmically addressing trade-offs. For example, if existing solutions predominantly tackle one of the facets, then there may be room for an algorithm that incorporates both and makes the trade-off controllable. 

Finally, we can interpret computing a variety of visualization candidates as a meta-problem within this trade-off space. That is, when we wish to offer multiple layout alternatives for a visual idiom (say, different cartograms of the same dataset), then we may want to ensure that, not only are these layouts good in the sense that they perform well in the various measures, but also that they are sufficiently distinct and cover the solution space (captured by the Pareto front) well.

Various approaches such as ``scagnostics'' \cite{wilkinson2005graph} and many others
\cite{behrisch2016magnostics,dasgupta2010pargnostics,shao2016guiding,bertini2011quality} find interesting views for a particular visual idiom. Though the various measures that underpin these approaches can be interpreted as a metric idiom, we observe that these techniques are orthogonal to trade-offs: rather than facilitating a trade-off in layout (``make the layout work for the data''), these methods fit data subsets to a given visual idiom (``make the data work for the visualization''). Yet, these types of approaches also sketch that measures and algorithmic thinking extend beyond layout questions, and we can aim to automate parts of the visual design itself, by determining informative views of data. A specific challenge to further automate visualization is to obtain quality metrics that overcome visual-idiom boundaries. Such measures may allow comparison between (possibly radically) different encodings and, if algorithmically actionable, be able to avoid a setup that requires generating and evaluating alternatives explicitly.

The trade-offs sketched above are different from hybrids in visual idioms, that seek to combine different encodings into a best-of-both-worlds solution. An example of such a hybrid is NodeTrix, combining matrix visualizations with node-link diagrams to represent a graph \cite{henry2007nodetrix}. Such hybrids mix visual idioms, contrasting trade-offs as intended above that seek to combine facets of quality within a visual idiom. For example, set visualizations may seek to optimize the Gestalt principle of grouping \cite{broek2024simplesets,collins2009bubble} or Tufte's minimal ink \cite{alper2011design,castermans2019short}, or offer a trade-off between these two extremes \cite{meulemans2013kelpfusion}.

\subsection{Hidden facets}\label{sec:hidden}

In the algorithmic perspective, an algorithm constructs a layout, by solving a computational problem derived from a metric idiom. Yet, a computational problem may allow for a variety of solutions that are equally good according to the computational problem, but that perform differently in the more general metric idiom. As a result, different algorithms for the same computational problem may give very different layouts, even if they perform similarly (or even optimally) when considered purely in context of the computational problem.

Yet, such different layouts may eventually lead to very different visualization quality overall, as interpreted in the metric idiom. That is, by design or by happenstance, the algorithms perform differently in facets not incorporated or constrained only loosely in the computational problem: we refer to such facets as the \emph{hidden facets} for an algorithm.

Algorithms can affect hidden facets either positively (effectively a proxy model) or negatively (effectively making an implicit trade-off). Especially for downstream validation, uncovering hidden facets and making them explicit is an important step to allow for fair and clear comparison of algorithms based on their properties.

\section{Integrating with Munzner's model}
\label{sec:integratemunzner}

To integrate the algorithmic perspective into Munzner's model \cite{munzner2009nested}, we inject a level that bears explicit consideration for establishing the metric idiom and computational problem (\autoref{fig:munzner}c) and refine the algorithm-design level. We follow Munzner's setup, describing for each step, its \emph{threats} (what can go wrong), its \emph{upstream validation} (how to assess whether the threats are appropriately dealt with before proceeding), and its \emph{downstream validation} (how to assess whether the threats were indeed appropriately dealt with, after later steps).

Note that we can interpret metric idiom design before algorithm design as how we are to evaluate algorithms downstream. Yet, the metric idiom also serves different purposes, in establishing adequacy and more generally allowing for comparison between visualizations from different origins. As such, we believe there is merit in treating this step separately. It also is an important conceptual distinction between \emph{what} to compute from \emph{how} to compute.

We observe that not every encoding or system may require explicit consideration of a metric idiom, or result in algorithmic challenges. The main question is whether the visual encoding leaves room for algorithmic choice: if it does, then defining quality measures is a necessary step for automating such choices and their evaluation.

\subsection{Metric idiom design}

Designing a metric idiom is outlined in \autoref{sec:algidiom}: establish legibility and correspondence facets, then define constraints and measures, and finally, cast this metric idiom into a computational problem. 

\paragraph{Threat.} 
The first threat is inadequacy: the metric idiom needs a good level of adequacy, to correspond well with human performance or preference. Otherwise, the metric idiom does not serve as a good predictor of visualization quality.

The second threat is unsolvability: we should be wary of overly convoluted computational problems, as they are cumbersome if not impossible to solve well. They may need disproportionate resources (time and/or memory) to achieve a reasonable solution.
We are assuming here, from the nested model, that the end goal is to build a system that produces high-quality visualizations. That is, we are interested in developing an algorithm that solves a computational problem derived from the metric idiom. For pure evaluation purposes, the permitted idiom complexity may be considerably higher; see also \autoref{sec:algidiom}.

It bears consideration here that, in the end, any metric idiom is a proxy of a very complex, non-standardized system---humans, influenced by capabilities, experience, preferences, and so forth. We need the metric idiom to mediate between this complexity and the ``simple'' formalism necessary for algorithmic study.

\paragraph{Upstream validation.}
Estimating the solvability is somewhat inherently tangled with the next step: developing the algorithm. Yet, keeping in mind that problem simplicity is a key factor here means that it can generally be advised to keep a concise model. On the other hand, additional constraints can help making a problem solvable, and one could eliminate constraints in the algorithm design step, if they prove too cumbersome to work with. That is, such problems one encounters already during the upstream stage of the algorithm design step. As Munzner \cite{munzner2009nested} also indicates, though the steps are presented as a nested model, realistically, one may go through the various steps several times before arriving at the eventual result. 

In comparison, assessing whether a metric idiom is (sufficiently) adequate ahead of time, is more important at this stage. It would be very inefficient to rely on algorithms design, system implementation and final downstream validation to establish that the constraints and measures chosen are inadequate for their purpose. 
There are various ways to estimate adequacy. For example, we may construct visualizations manually according to the principles set by the metric idiom, or quickly implement a brute-force solution to solve small instances and evaluate its results. We can further use it to assess existing visualizations---if done right, a metric idiom is independent of the construction mechanism after all---or ground them in existing guidelines and literature.

A more theoretic approach is to apply worst-case thinking, devising counterexamples: datasets in which the metric idiom ``clearly'' gives the wrong prediction, thereby highlighting missing constraints or flaws in the measure. However, as also discussed before, we can likely always find some mispredictions. The question therefore lies in how degenerate or unrealistic the hypothetical dataset has to become, to reveal the flaw, and the severity of the flaw itself.

\paragraph{Downstream validation.}
Solvability at this point should already be addressed, as this matches the downstream validation of the algorithm-design step (see \autoref{sec:nested_algorithm}). So, what remains to establish is the adequacy of the metric idiom. That is, we need to correlate human performance or preference to quality as measured by the metric idiom.

The typical case would be to use the developed algorithm to construct actual visualizations, and use these in a controlled experiment involving human participants. The advantage of doing so, is that we gain information about how well the algorithm actually performs, because it achieves a certain (presumably, high) level of quality according to the measures as established in the algorithm-design step. A possible drawback is that it may be hard to disentangle the algorithmic result from the idiom's adequacy. That is, solutions are often not unique and leave considerable room for other hidden factors of quality, that are implicitly affected by the algorithm.

But as also alluded to above, a metric idiom's purpose is evaluation and should not depend on an algorithm. That is, one should be able to take any visualization in the given visual idiom (appropriately encoded) and evaluate the measures and constraints. The advantage of using independently constructed visualizations is that one may get a better coverage of good visualizations that may not necessarily structurally make the same decisions based on hidden factors. That is, we can also aim to more broadly and directly capture adequacy of a metric idiom, by using a varied set of sources for the stimuli.

Often, algorithmic visualization papers do consider deriving a computational problem, or at least discuss metrics for evaluation, without a formally establishing their adequacy. This is typically necessary both to keep the focus of the paper and to stay with one's expertise: generally establishing adequacy is far from a trivial matter, or an afterthought. The upstream validation steps may help in convincing a reader of some basic level of adequacy of the modeled problem. 

A downstream counterpart is then often a \emph{qualitative image analysis}, as also present at the encoding technique design level in Munzner's model \cite{munzner2009nested}. That is, one uses algorithms to generate visualizations according to the principles of the metric idiom (or more directly, of its computational problem), to then discuss the strengths and weaknesses of the resulting visualizations. Such practice serves to validate that the choices made in developing the metric idiom, computational problem and algorithm indeed lead to high-quality visualizations or uncover weaknesses that merit further algorithmic consideration. Albeit often subjectively assessed, it allows a reader to either agree or disagree. Such a discussion can also serve to uncover hidden facets, or establish relations between facets that were otherwise reduced for the sake of the computational problem.

Another step in downstream validation here is to evaluate beyond the computational problem itself. That is, to evaluate using the multifaceted metric idiom instead: to find correlations or trade-offs between facets in practical application. This may help to, for example, justify a simple computational problem as it relates well to quality as captured according to other facets. But it may also encompass comparison to algorithms for another computational problem, in trying to uncover which algorithm (or even, computational problem) eventually performs ``best'' across the various facets of the metric idiom.

\subsection{Algorithm design}\label{sec:nested_algorithm}

With a computational problem at hand, the algorithm design step can focus on solving that particular problem well. This is relatively standard practice across algorithms theory \cite{cormen2022introduction} and engineering \cite{mendling2025methodology}, even if approaches vary between subfields. As such, we do not go into detail here and discuss considerations only briefly.

\paragraph{Threat.}
A first threat is that an algorithm requires too much resources (typically, time or memory) to meet the necessary efficiency requirements. Though note that, depending on the final application, efficiency requirements may vary greatly: an interactive visual-analytics system may need to be able to generate visualizations at (near-)interactive speed. Yet, an algorithm for constructing an explanatory data visualization to support an article or to feature in an atlas may often take considerably more time.

A second threat is that the algorithm does not achieve sufficient quality: it risks violating constraints, or does not achieve a high value in the optimization objective, compared to the optimal solution. Note that the first would in fact be an invalid algorithm: it does not actually solve the computational problem. The second would suggest that it applies heuristics that are not necessarily effective.

Note that both considerations are also in Munzner's exposition~\cite{munzner2009nested}, though the second threat is mostly restricted to ``correctness''.

\paragraph{Upstream validation.}
Before implementing an algorithm, we can already estimate its resource requirements using standard asymptotic analysis of running time, memory or other resources. For assessing the quality of an algorithm, we can aim to prove that it achieves an optimal solution, or achieves a certain approximation ratio.

\paragraph{Downstream validation.}
With an implementation available, we can measure actual resource usage and quality (according to the optimization objective). Resources can then be compared to the requirements of the application, or to those used by other algorithms. Quality can be compared to the optimal solution (obtained by a brute-force solver, for example) or to solutions obtained by other algorithms. Such algorithms could in principle solve different computational problems, or we could even use other sources such as manually constructed visualizations for comparison. The primary purpose is to understand how well the developed algorithm solves the designed computational problem.

\section{On algorithmic, convergent and divergent thinking}
\label{sec:convergent-divergent}

The purpose of this paper lies with establishing a model for the interaction between visualization and algorithms research, not with developing a model of how to solve algorithmic challenges. Yet, to foster understanding and collaboration across these fields, it may be beneficial to reflect on standard strategies of solving complex problems.

Algorithmic work tends to focus on well-defined problems: ``given an input, how can we compute an output that minimizes this well-defined function?''. Yet, as established above, computational challenges encountered in visualization (and many other fields) are often ill-defined: there is no universally agreed function that captures ``good''. 

The algorithmic perspective aims to provide structures to bridge this gap. That is, it focuses on the \emph{problem construction} phase of dealing with ill-defined problems. As an extensive exposition of the cognitive processes is out of scope here, we refer to the work of Wigert et al.~\cite{wigert2024utility} and the references therein for elaboration.

But even when we have constructed a metric idiom and derived a computational problem, solution strategies may vary between fields.
Literature suggests that such approaches generally take the form of (alternating) divergent and convergent phases \cite{guilford1956structure}: divergent thinking to collect ideas and alternatives, followed by convergent thinking to prioritize and select to obtain a solution. Divergent thinking broadens the scope to look for possibilities. Convergent thinking then narrows the broadened scope down to an eventual solution. Note that these phases are also reflected in the basic design cycle for visualization, described by Van Wijk \cite{wijk2006views}.

However, in theoretical algorithms research, this process \emph{seems} different. To solve a complex problem, one may start with trying to simplify the problem: impose restrictions on the input, or to simplify the goal by adding or removing constraints on the solution. The goal is then to take the lessons learned in the simplified setting, to subsequently generalize the solution to solve the more complex problem. That is, the process starts with narrowing down (which may feel convergent), followed by broadening the scope (which may appear divergent).

\begin{figure}[t]
    \centering
    \includegraphics[alt={Two diagrams labeled (a) and (b) show sets of arrows, all roughly pointing towards the right. Diagram (a) shows a set of red arrows labeled ``diverge'' on the left with a common starting point, but different end points and a set of blue arrows labeled ``converge'' on the right with different starting points but a common end point. Each end of a red arrow matches with the start of a blue arrow. The result is a diamond-shaped outline of the diagram. Diagram (b) inverses this picture: the blue arrows are now on the left and are labeled ``simplified''; the red arrows are now on the left are are labeled ``generalize''. The common end point of the blue arrows now matches the common start point of the red arrows. The result is a bow-tie-shaped outline of the diagram.},page=1]{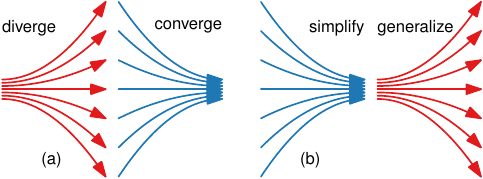}
    \caption{(a) The standard problem-solving approach of divergent, then convergent thinking. (b) The seemingly opposite algorithmic problem-solving model of simplification, then generalization.}
    \label{fig:divergent-convergent}
\end{figure}

Schematically, this looks opposite: see \autoref{fig:divergent-convergent}. Yet, in the end, the simplification phase \emph{is} a form of divergent thinking, as it consists of trying out different directions of how to solve the complex problem; the generalization phase \emph{is} a form of convergent thinking, as it consists of selecting which aspects of the simplified problem can be useful for the more complex problem.

While simplification can be interpreted as adding requirements in the basic design cycle \cite{wijk2006views}, such requirements are not on the eventual result but rather on the context or input. The simplification process may feel to be moving further and further from the problem that was to be solved initially. Therefore, especially in the first phase, the different approaches may lead to confusion or even frustration if the process is not clear in a collaboration. It bears consideration that the eventual generalization phase is to come.

\section{Examples}
\label{sec:examples}

Below, we briefly discuss various visualization types, interpreting existing literature in context of the proposed algorithmic perspective.

\subsection{Matrix ordering: the role of computational problems}

Visualizing a graph via its adjacency matrix is a standard technique, but it relies crucially on choosing a good ordering of the vertices (rows and columns) such that meaningful patterns emerge \cite{behrisch2016matrix}. Various quality measures and algorithms for matrix orderings exist \cite{behrisch2016matrix}, yet their origins and thereby their more abstract concept of ``good ordering'' varies \cite{liiv2010seriation}. For example, bandwidth and other distance-to-diagonal measures may not correlate well to patterns: quantifying and optimizing similarity between neighborhoods is more adequate in making patterns emerge \cite{beusekom2021simultaneous}. Understanding the types of computational problems and how they relate to quality facets, or at least how they embody the concept of ``good'', is therefore crucial in selecting appropriate algorithms for ordering matrices for visualization purposes.

Bach et al.~\cite{bach2015small} describe an interactive system for collections of graphs visualized through matrices. One particular feature is that all or a subset of matrices can be ordered to find ``topological clusters'' across all graphs. The purpose of this particular work is not algorithmic, and the computational problem here goes undefined. Reverse-engineering, Van Beusekom et al.~\cite{beusekom2021simultaneous} state that the algorithm by Bach et al.\ effectively sums all matrices to apply a standard reordering algorithm. However, as Van Beusekom et al.~argue and indeed show, a ``good'' ordering for multiple matrices does not necessarily match to a good ordering of their sum: instead, such a simultaneous ordering should achieve a high quality by summing the resulting qualities. They then show how to modify existing ordering algorithms to operate on collections of matrices instead.
That is, by explicitly considering the computational problem, one may spot logical flaws in how algorithms are applied.

\subsection{Grid maps: leveraging the metric idiom}

Grid maps (spatially arranged small multiples) offer a complex case where the multifaceted nature of ``maintaining geography well'' arises. Through computational experiments, it was established that minimizing the displacement between region and cell centroid offers a simple but good proxy for other facets such as relative directions and topology \cite{eppstein2015improved}, at least in cases where the target grid has (approximately) as many cells as the dataset has regions. That is, displacement acts as a good proxy model for the various other facets.
The primary contribution perhaps lies in the algorithm design level, yet also discusses modeling the problem and comparing different computational problems. In our model, this lies in the metric idiom design level.

Meulemans et al.~\cite{meulemans2016small} extended the metric idiom, exploring relations between facets in more depth. In this extended idiom, the results by Eppstein et al.~\cite{eppstein2015improved} were validated. However, it also revealed that considerations of shape and connectivity change as the grid size grows to accommodate whitespace and thereby allow more flexibility in assigning regions to cells.
As such, this research studies the metric idiom design, yet leveraging algorithmic contributions to support the study in downstream validation.

These insights eventually led to a pipeline for computing grid maps \cite{meulemans2020simple}: each of the three steps solves a separate computational problem, yet they together achieve a solution for a complex problem, specifically by setting constraints such as general contiguity on the overall pipeline. 
The downside of such a stepwise approach is that the overall computational problem is less crisply defined. Nonetheless, the metric idiom explored by Meulemans et al.~\cite{meulemans2016small} allows for quantifying quality of the results. Recently, this idiom and derivations thereof were also applied to compare results between algorithms explicitly \cite{meulemans2026newpipeline}.
In both, qualitative image analysis is also applied to uncover limitations to the metrics used for evaluation.

\subsection{Overlap removal: uncovering opportunities}
\label{sec:overlapremoval}

Symbols often overlap, when their coordinates are prescribed (for example, in symbol maps or scatter plots) or generated by algorithms that are agnostic to symbol size (such as various graph-drawing algorithms). To improve legibility, one may consider displacing elements to reduce overlap or remove it altogether. 
Van Garderen et al.~\cite{garderen2017minimum} studied the computational problem of minimizing displacement of square or rectangular symbols, constrained such that the symbol centers maintain their x- and y-orders: this problem is NP-hard in general, and as such they develop heuristics.

Subsequently, it was observed that changing the symbols from squares to diamonds (or alternatively, rotating the order constraints) in fact yields a considerably simpler problem, one that can be solved optimally though linear programming \cite{meulemans2019efficient}. That is, by formally defining the computational problem, one may also observe that even slight changes to this computational problem may in fact yield very different algorithmic results. Of course, the adequacy of the model may be lower, depending on the specific use case: it then lies with an eventual visualization developer or designer to decide whether the distinction in problems is relevant to warrant using one algorithm or another. Yet, by elucidating the computational problem, such choices become possible in striking a balance between adequacy and solvability (\autoref{sec:adequacy}).

\subsection{Thematic maps: simplify and generalize}

It is often infeasible to cast the whole complexity of a metric idiom into a single, encompassing computational problem. Nonetheless, a formal algorithmic approach can be helpful in achieving high-quality visualizations, effectively demonstrating the simplify-generalize process as outlined in \autoref{sec:convergent-divergent}. We observe this, for example, in two thematic mapping algorithms: necklace maps \cite{speckmann2010necklace} and flow maps \cite{buchin2011flow}.

In both cases, the computational problem is designed in such a way (``simplified'') as to allow algorithms with provable guarantees, on the core structure of the eventual layout. Yet, both approaches explicitly acknowledge that the results do not fulfill all facets of quality, and thus use postprocessing to ``generalize'' the solution in considering further quality facets, while ensuring that the core structure remains sound. Also here, qualitative image analysis is used, not to (just) assess the visual encoding itself, but rather to discuss strengths and weaknesses of the modeled computational problem.

\section{Implications of the algorithmic perspective}
\label{sec:implications}

The purpose of this algorithmic perspective is to bring computational modeling to the foreground of approaching visualization algorithms. The metric idiom serves as a quantifiable exploration of visualization quality, and as a basis to derive computational problems. We need such formalism to be able to: (1) discuss priorities, constraints and limitations of algorithms and techniques; (2) to evaluate and compare algorithms in a clear manner; and (3) to relate computational problems and understand trade-offs and proxies in a quantifiable way.

Akin to how reporting a user evaluation is expected to clearly defend its design choices and outline its limitations, visualization research that introduces new algorithms or applies existing algorithms outside their original context should: clearly outline concepts of quality (metric idiom); defend the suitability of the computational problem underpinning the introduced or applied algorithm; and discuss limitations that these have on the reported results. And while we focused on layout algorithms here, these concerns are valid in other contexts as well, as soon as algorithms are used to make decisions that impact the eventual visualizations.

We should be wary to wrongfully conclude that the algorithmic perspective is (just) about formalism for guarantees. Heuristics clearly have value due to the multifaceted, and often complex nature of layout problems. The point is to specify problems---or at least measurable facets of quality---as to allow for comparison, and understanding the limitations that the computational problem or the algorithm itself cause in further downstream evaluation.

The complexity of computational problems in visualization is not a reason to shy away from specifying quality considerations: it does not take away the need for automating visualization construction and understanding the algorithms and systems that support such.

In an ideal world, human-centered evaluation is used to establish adequacy of metric idioms and computational problems. With adequacy established, algorithmic visualization research can focus on solving such problems well, and will not need user evaluation. Yet, establishing adequacy is far from trivial and indeed often beyond the scope (and possibly, expertise) for algorithmic research. Asking for user evaluation of algorithmic contributions seems an oddity: it is optimizing a well-specified problem after all. The question should rather revolve around the decisions made in developing its computational problem (adequacy). While  evaluation with humans can give very actionable results (``in this context, this system outperforms that system for such tasks''), we should be aware this practice results in inseparable stacks \cite{meyer2015nested}: it does not actually test the algorithm, but rather whether, together with the upstream decisions, it achieves a certain level of performance. As a result, it prevents separate algorithmic improvements from carrying over into these actionable results. Indeed, neither in our model nor in Munzner's model \cite{munzner2009nested} are user studies and alike a downstream validation method at the algorithm level.

In the end, the main point of the algorithmic perspective is to clearly separate concerns: \emph{what} we should be computing, versus \emph{how} do we compute a good result? While there is abundant research focusing on the latter question, the former seems more scarce. It bears consideration how we can promote and value the former question more clearly in algorithmic visualization research, understanding that actually establishing adequacy is typically beyond scope when developing algorithms. 

\acknowledgments{
The author would like to thank Bettina Speckmann for inspiring discussions on the interplay between algorithms research and visualization research, and Helen Purchase for exchanging thoughts and diagrams on the correspondence between stimulus metric measures and human performance measures. 
W. Meulemans is partially supported by the Dutch Research Council (NWO) under project number VI.Vidi.223.137.}

\bibliographystyle{abbrv-doi-hyperref}

\bibliography{references}

@article{munzner2009nested,
  title={A nested model for visualization design and validation},
  author={Munzner, Tamara},
  journal={IEEE Transactions on Visualization and Computer Graphics},
  volume={15},
  number={6},
  pages={921--928},
  year={2009},
  doi={10.1109/TVCG.2009.111}
}

@book{munzner2014visualization,
author = {T. Munzner},
title = {Visualization analysis and design},
publisher = {AK Peters/CRC Press},
year = {2014}
}

@article{chen2015what,
  title={What may visualization processes optimize?},
  author={Chen, Min and Golan, Amos},
  journal={IEEE Transactions on Visualization and Computer Graphics},
  volume={22},
  number={12},
  pages={2619--2632},
  year={2016},
  doi={10.1109/TVCG.2015.2513410}
}

@article{meyer2015nested,
  title={The nested blocks and guidelines model},
  author={Meyer, Miriah and Sedlmair, Michael and Quinan, P Samuel and Munzner, Tamara},
  journal={Information Visualization},
  volume={14},
  number={3},
  pages={234--249},
  year={2015},
  doi={10.1177/1473871613510429}
}

@article{wijk2006views,
  title={Views on visualization},
  author={Van Wijk, Jarke J},
  journal={IEEE Transactions on Visualization and Computer Graphics},
  volume={12},
  number={4},
  pages={421--432},
  year={2006},
  doi={10.1109/TVCG.2006.80}
}

@article{wigert2024utility,
  title={The utility of divergent and convergent thinking in the problem construction processes during creative problem-solving},
  author={Wigert, Benjamin G and Murugavel, Vignesh R and Reiter-Palmon, Roni},
  journal={Psychology of Aesthetics, Creativity, and the Arts},
  volume={18},
  number={5},
  pages={858--868},
  year={2024},
  doi={10.1037/aca0000513}
}

@article{guilford1956structure,
  title={The structure of intellect},
  author={Guilford, J. P.},
  journal={Psychological bulletin},
  volume={53},
  number={4},
  pages={267--293},
  year={1956},
  doi={10.1037/h0040755}
}

@article{eppstein2015improved,
  title={Improved grid map layout by point set matching},
  author={Eppstein, David and van Kreveld, Marc and Speckmann, Bettina and Staals, Frank},
  journal={International Journal of Computational Geometry \& Applications},
  volume={25},
  number={02},
  pages={101--122},
  year={2015},
  doi={10.1142/S0218195915500077}
}

@article{meulemans2016small,
  title={Small multiples with gaps},
  author={Meulemans, Wouter and Dykes, Jason and Slingsby, Aidan and Turkay, Cagatay and Wood, Jo},
  journal={IEEE Transactions on Visualization and Computer Graphics},
  volume={23},
  number={1},
  pages={381--390},
  year={2017},
  doi={10.1109/TVCG.2016.2598542}
}

@article{meulemans2020simple,
  title={A simple pipeline for coherent grid maps},
  author={Meulemans, Wouter and Sondag, Max and Speckmann, Bettina},
  journal={IEEE Transactions on Visualization and Computer Graphics},
  volume={27},
  number={2},
  pages={1236--1246},
  year={2021},
  doi={10.1109/TVCG.2020.3028953}
}

@article{meulemans2026newpipeline,
title={A Simple Grid-Maps Pipeline: Restructured, Accelerated and Upgraded},
author={Meulemans, Wouter},
journal={Computer Graphics Forum},
year={2026},
volume={45},
number={3},
note={Forthcoming},
doi={10.1111/cgf.70467}
}

@article{meulemans2019efficient,
  title={Efficient optimal overlap removal: Algorithms and experiments},
  author={Meulemans, Wouter},
  journal={Computer Graphics Forum},
  volume={38},
  number={3},
  pages={713--723},
  year={2019},
  doi={10.1111/cgf.13722}
}

@article{garderen2017minimum,
  title={Minimum-displacement overlap removal for geo-referenced data visualization},
  author={van Garderen, Mereke and Pampel, Barbara and Nocaj, Arlind and Brandes, Ulrik},
  journal={Computer Graphics Forum},
  volume={36},
  number={3},
  pages={423--433},
  year={2017},
  doi={10.1111/cgf.13199}
}

@article{mendling2025methodology,
  title={Methodology of algorithm engineering},
  author={Mendling, Jan and Leopold, Henrik and Meyerhenke, Henning and Depaire, Beno{\^\i}t},
  journal={ACM Computing Surveys},
  volume={58},
  number={4},
  pages={1--38},
  year={2025},
  doi={10.1145/3769071}
}

@book{cormen2022introduction,
  title={Introduction to algorithms},
  author={Cormen, Thomas H and Leiserson, Charles E and Rivest, Ronald L and Stein, Clifford},
  year={2022},
  publisher={MIT Press}
}

@article{lehmann2015study,
  title={A study on quality metrics vs. human perception: Can visual measures help us to filter visualizations of interest?},
  author={Lehmann, Dirk J and Hundt, Sebastian and Theisel, Holger},
  journal={Information Technology},
  volume={57},
  number={1},
  pages={11--21},
  year={2015},
  doi={10.1515/itit-2014-1070}
}

@inproceedings{purchase1995validating,
  title={Validating graph drawing aesthetics},
  author={Purchase, Helen C and Cohen, Robert F and James, Murray},
  booktitle={Proceedings of the International Symposium on Graph Drawing},
  pages={435--446},
  year={1995},
  series={LNCS 1027},
  doi={10.1007/BFb0021827}
}

@article{seldmair2015datadriven,
author = {Sedlmair, M. and Aupetit, M.},
title = {Data-driven Evaluation of Visual Quality Measures},
journal = {Computer Graphics Forum},
volume = {34},
number = {3},
pages = {201--210},
doi = {https://doi.org/10.1111/cgf.12632},
year = {2015}
}

@article{speckmann2010necklace,
  title={Necklace maps},
  author={Speckmann, Bettina and Verbeek, Kevin},
  journal={IEEE Transactions on Visualization and Computer Graphics},
  volume={16},
  number={6},
  pages={881--889},
  year={2010},
  doi={10.1109/TVCG.2010.180}
}

@article{buchin2011flow,
  title={Flow map layout via spiral trees},
  author={Buchin, Kevin and Speckmann, Bettina and Verbeek, Kevin},
  journal={IEEE Transactions on Visualization and Computer Graphics},
  volume={17},
  number={12},
  pages={2536--2544},
  year={2011},
  doi={10.1109/TVCG.2011.202}
}

@article{cowley2022framework,
  title={A framework for rigorous evaluation of human performance in human and machine learning comparison studies},
  author={Cowley, Hannah P and Natter, Mandy and Gray-Roncal, Karla and Rhodes, Rebecca E and Johnson, Erik C and Drenkow, Nathan and Shead, Timothy M and Chance, Frances S and Wester, Brock and Gray-Roncal, William},
  journal={Scientific Reports},
  volume={12},
  number={1},
  pages={Art. 5444},
  year={2022},
  doi={10.1038/s41598-022-08078-3}
}

@article{beusekom2021simultaneous,
  title={Simultaneous matrix orderings for graph collections},
  author={van Beusekom, Nathan and Meulemans, Wouter and Speckmann, Bettina},
  journal={IEEE Transactions on Visualization and Computer Graphics},
  volume={28},
  number={1},
  pages={1--10},
  year={2022},
  doi={10.1109/TVCG.2021.3114773}
}

@article{behrisch2016matrix,
  title={Matrix reordering methods for table and network visualization},
  author={Behrisch, Michael and Bach, Benjamin and Henry Riche, Nathalie and Schreck, Tobias and Fekete, Jean-Daniel},
  journal={Computer Graphics Forum},
  volume={35},
  number={3},
  pages={693--716},
  year={2016},
  doi={10.1111/cgf.12935}
}

@article{liiv2010seriation,
  title={Seriation and matrix reordering methods: An historical overview},
  author={Liiv, Innar},
  journal={Statistical Analysis and Data Mining: The ASA Data Science Journal},
  volume={3},
  number={2},
  pages={70--91},
  year={2010},
  doi={10.1002/sam.10071}
}

@article{bach2015small,
  title={Small {MultiPiles}: Piling time to explore temporal patterns in dynamic networks},
  author={Bach, Benjamin and Henry-Riche, Nathalie and Dwyer, Tim and Madhyastha, Tara and Fekete, J-D and Grabowski, Thomas},
  journal={Computer Graphics Forum},
  volume={34},
  number={3},
  pages={31--40},
  year={2015},
  doi={10.1111/cgf.12615}
}

@inproceedings{cabouat2026readability,
  title={Readability as a multi-measure construct in data visualization},
  author={Cabouat, Anne-Flore and Huron, Samuel and Isenberg, Tobias and Isenberg, Petra},
  booktitle={Proceedings of the CHI 2026 STAR Workshop -- Science and Technology for Augmenting Reading},
  pages={1--5},
  url={https://hal.science/hal-05548028v1},
  year={2026}
}

@article{henry2007nodetrix,
  title={Node{T}rix: a hybrid visualization of social networks},
  author={Henry, Nathalie and Fekete, Jean-Daniel and McGuffin, Michael J},
  journal={IEEE Transactions on Visualization and Computer Graphics},
  volume={13},
  number={6},
  pages={1302--1309},
  year={2007},
  doi={10.1109/TVCG.2007.70582}
}

@inproceedings{wilkinson2005graph,
  title={Graph-theoretic scagnostics},
  author={Wilkinson, Leland and Anand, Anushka and Grossman, Robert},
  booktitle={Proceedings of the IEEE Symposium on Information Visualization},
  pages={21--21},
  year={2005},
  doi={10.1109/INFVIS.2005.1532142}
}

@article{shao2016guiding,
  title={Guiding the exploration of scatter plot data using motif-based interest measures},
  author={Shao, Lin and Schleicher, Timo and Behrisch, Michael and Schreck, Tobias and Sipiran, Ivan and Keim, Daniel A},
  journal={Journal of Visual Languages \& Computing},
  volume={36},
  pages={1--12},
  year={2016},
  doi={10.1016/j.jvlc.2016.07.003}
}

@article{dasgupta2010pargnostics,
  title={Pargnostics: Screen-space metrics for parallel coordinates},
  author={Dasgupta, Aritra and Kosara, Robert},
  journal={IEEE Transactions on Visualization and Computer Graphics},
  volume={16},
  number={6},
  pages={1017--1026},
  year={2010},
  doi={10.1109/TVCG.2010.184}
}

@article{behrisch2016magnostics,
  title={Magnostics: Image-based search of interesting matrix views for guided network exploration},
  author={Behrisch, Michael and Bach, Benjamin and Hund, Michael and Delz, Michael and Von R{\"u}den, Laura and Fekete, Jean-Daniel and Schreck, Tobias},
  journal={IEEE Transactions on Visualization and Computer Graphics},
  volume={23},
  number={1},
  pages={31--40},
  year={2017},
  doi={10.1109/TVCG.2016.2598467}
}

@book{chen2020foundations,
  title={Foundations of data visualization},
  author={Chen, Min and Hauser, Helwig and Rheingans, Penny and Scheuermann, Gerik},
  year={2020},
  publisher={Springer},
  doi={10.1007/978-3-030-34444-3}
}

@article{kindlmann2014algebraic,
  title={An algebraic process for visualization design},
  author={Kindlmann, Gordon and Scheidegger, Carlos},
  journal={IEEE Transactions on Visualization and Computer Graphics},
  volume={20},
  number={12},
  pages={2181--2190},
  year={2014},
  doi={10.1109/TVCG.2014.2346325}
}

@inproceedings{wijk2005value,
  title={The value of visualization},
  author={Van Wijk, Jarke J},
  booktitle={Proceedings of the IEEE Conference on Visualization},
  pages={79--86},
  year={2005},
  doi={10.1109/VISUAL.2005.1532781}
}

@article{cabello2010algorithmic,
  title={Algorithmic aspects of proportional symbol maps},
  author={Cabello, Sergio and Haverkort, Herman and Van Kreveld, Marc and Speckmann, Bettina},
  journal={Algorithmica},
  volume={58},
  number={3},
  pages={543--565},
  year={2010},
  doi={10.1007/s00453-009-9281-8}
}

@article{giovannangeli2023guaranteed,
  title={Guaranteed visibility in scatterplots with tolerance},
  author={Giovannangeli, Loann and Lalanne, Frederic and Giot, Romain and Bourqui, Romain},
  journal={IEEE Transactions on Visualization and Computer Graphics},
  volume={30},
  number={1},
  pages={792--802},
  year={2024},
  doi={10.1109/TVCG.2023.3326596}
}

@article{gartner2025optimizing,
  title={Optimizing symbol visibility through displacement},
  author={G{\"a}rtner, Bernd and Kalani, Vishwas and Reddy, Meghana M and Meulemans, Wouter and Speckmann, Bettina and Stojakovi{\'c}, Milo{\v{s}}},
  journal={Applied Mathematics and Computation},
  volume={505},
  pages={Art. 129529},
  year={2025},
  doi={10.1016/j.amc.2025.129529}
}

@article{chen2017pathways,
  title={Pathways for theoretical advances in visualization},
  author={Chen, Min and Grinstein, Georges and Johnson, Chris R and Kennedy, Jessie and Tory, Melanie},
  journal={IEEE Computer Graphics and Applications},
  volume={37},
  number={4},
  pages={103--112},
  year={2017},
  doi={10.1109/MCG.2017.3271463}
}

@article{broek2024simplesets,
  title={Simple{S}ets: Capturing Categorical Point Patterns with Simple Shapes},
  author={van den Broek, Steven and Meulemans, Wouter and Speckmann, Bettina},
  journal={IEEE Transactions on Visualization and Computer Graphics},
  volume={31},
  number={1},
  pages={262--271},
  year={2025},
  doi={10.1109/TVCG.2024.3456168}
}

@article{alper2011design,
  title={Design study of {LineSets}, a novel set visualization technique},
  author={Alper, Basak and Riche, Nathalie and Ramos, Gonzalo and Czerwinski, Mary},
  journal={IEEE Transactions on Visualization and Computer Graphics},
  volume={17},
  number={12},
  pages={2259--2267},
  year={2011},
  doi={10.1109/TVCG.2011.186}
}

@article{collins2009bubble,
  title={Bubble {S}ets: Revealing set relations with isocontours over existing visualizations},
  author={Collins, Christopher and Penn, Gerald and Carpendale, Sheelagh},
  journal={IEEE Transactions on Visualization and Computer Graphics},
  volume={15},
  number={6},
  pages={1009--1016},
  year={2009},
  doi={10.1109/TVCG.2009.122}
}

@article{castermans2019short,
  title={Short plane supports for spatial hypergraphs},
  author={Castermans, Thom and Van Garderen, Mereke and Meulemans, Wouter and N{\"o}llenburg, Martin and Yuan, Xiaoru},
  journal={Journal of Graph Algorithms and Applications},
  volume={23},
  number={3},
  pages={463--498},
  year={2019},
  doi={10.7155/jgaa.00499}
}

@article{meulemans2013kelpfusion,
  title={Kelp{F}usion: A hybrid set visualization technique},
  author={Meulemans, Wouter and Riche, Nathalie Henry and Speckmann, Bettina and Alper, Basak and Dwyer, Tim},
  journal={IEEE Transactions on Visualization and Computer Graphics},
  volume={19},
  number={11},
  pages={1846--1858},
  year={2013},
  doi={10.1109/TVCG.2013.76}
}

@article{zhao2020preserving,
  title={Preserving minority structures in graph sampling},
  author={Zhao, Ying and Jiang, Haojin and Chen, Qi'an and Qin, Yaqi and Xie, Huixuan and Wu, Yitao and Liu, Shixia and Zhou, Zhiguang and Xia, Jiazhi and Zhou, Fangfang},
  journal={IEEE Transactions on Visualization and Computer Graphics},
  volume={27},
  number={2},
  pages={1698--1708},
  year={2021},
  doi={10.1109/TVCG.2020.3030428}
}

@article{behrisch2018quality,
  title={Quality metrics for information visualization},
  author={Behrisch, Michael and Blumenschein, Michael and Kim, Nam Wook and Shao, Lin and El-Assady, Mennatallah and Fuchs, Johannes and Seebacher, Daniel and Diehl, Alexandra and Brandes, Ulrik and Pfister, Hanspeter and others},
  journal={Computer Graphics Forum},
  volume={37},
  number={3},
  pages={625--662},
  year={2018},
  doi={10.1111/cgf.13446}
}

@article{bertini2011quality,
  title={Quality metrics in high-dimensional data visualization: An overview and systematization},
  author={Bertini, Enrico and Tatu, Andrada and Keim, Daniel},
  journal={IEEE Transactions on Visualization and Computer Graphics},
  volume={17},
  number={12},
  pages={2203--2212},
  year={2011},
  doi={ 10.1109/TVCG.2011.229}
}

@article{moritz2019draco,
author={Moritz, D. and Wang, C. and Nelson, G. L. and Lin, H. and Smith, A. M. and Howe, B. and Heer, J.},
title={Formalizing visualization design knowledge as constraints: Actionable and extensible models in {D}raco},
journal={IEEE Transactions on Visualization and Computer Graphics},
volume={25},
number={1},
pages={438--448},
year={2019},
doi={10.1109/TVCG.2018.2865240}
}

@article{satyanarayan2016vega,
  title={{Vega-Lite}: A grammar of interactive graphics},
  author={Satyanarayan, Arvind and Moritz, Dominik and Wongsuphasawat, Kanit and Heer, Jeffrey},
  journal={IEEE Transactions on Visualization and Computer Graphics},
  volume={23},
  number={1},
  pages={341--350},
  year={2017},
  doi={10.1109/TVCG.2016.2599030}
}

\end{document}